\documentclass[a4paper, 10pt, twocolumn]{article} 
\usepackage[british]{babel}
\usepackage{microtype}
\usepackage{amsmath,amsfonts,amsthm}
\usepackage[svgnames,table]{xcolor}
\definecolor{GammaDarkGreen}{rgb}{0.113,0.278,0.2745}
\definecolor{GammaCharcoal}{rgb}{0.0706, 0.0706, 0.0706}
\definecolor{GammaRed}{HTML}{B71325}
\definecolor{GammaGreen}{HTML}{004846}
\usepackage{booktabs}
\usepackage{lastpage}
\usepackage{graphicx}
\usepackage{enumitem}
\setlist{leftmargin=*,itemsep=0.4ex,parsep=0ex,topsep=0.8ex,partopsep=0ex}
\setlist[itemize]{label=\color{GammaRed}$\triangleright$}
\usepackage{sectsty}
\chapterfont{\normalfont\Large\color{GammaRed}}
\sectionfont{\normalfont\Large\color{GammaRed}}
\subsectionfont{\normalfont\large\color{GammaCharcoal}}
\subsubsectionfont{\normalfont\normalsize\color{GammaGreen}}

\usepackage{subfig}

\usepackage{caption}
\DeclareCaptionLabelFormat{abbrevfig}{\color{GammaRed}Fig.\ #2}
\DeclareCaptionLabelFormat{abbrevtab}{\color{GammaRed}Tab.\ #2}
\usepackage[pdftex]{hyperref}
\pdftrailerid{}
\hypersetup{colorlinks=true,linkcolor=GammaDarkGreen,citecolor=GammaDarkGreen,filecolor=GammaDarkGreen,urlcolor=GammaDarkGreen}

\addto\extrasenglish{%
}

\usepackage{geometry}
\usepackage[utf8]{inputenc}
\usepackage[T1]{fontenc}
\IfFileExists{nimbusmononarrow.sty}{\usepackage{nimbusmononarrow}}{}
\usepackage{tgheros}
\usepackage{sansmath}
\usepackage{fancyhdr}
\usepackage{titling}

\newcommand{\authorstyle}[1]{{\large\color{GammaCharcoal}#1}}
\newcommand{\institution}[1]{{\footnotesize\itshape\color{GammaDarkGreen}#1\/}}

\pretitle{%
    \noindent
    \begin{minipage}[t]{0.63\textwidth}\vspace{20pt}
        \LARGE\color{GammaRed}
}

\posttitle{%
    \end{minipage}%
    \hfill
    \begin{minipage}[t]{0.33\textwidth}\vspace{-50pt}
        \raggedleft
        \includegraphics[width=\linewidth]{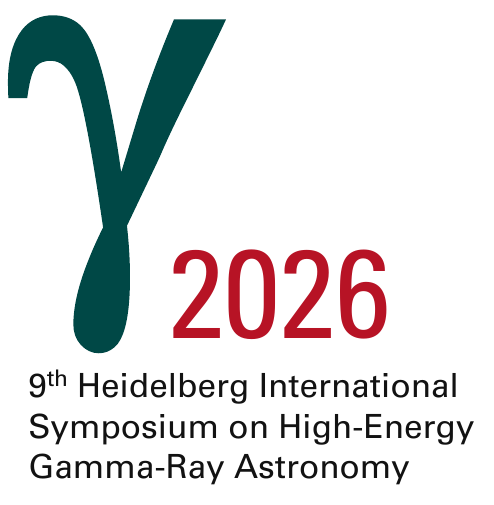}
    \end{minipage}%
}

\preauthor{\color{GammaDarkGreen}}
\postauthor{}
\predate{\begin{flushleft}\large\color{GammaRed}}
\postdate{\par\end{flushleft}}

\usepackage{lettrine}
\usepackage{xstring}
\newcommand{\initial}[1]{\lettrine[lines=2,findent=4pt,nindent=0pt]{\color{GammaRed}\emph{#1}}{}}
\newcommand{\lettrineabstract}[1]{\StrLeft{#1}{1}[\firstletter]\initial{\firstletter}\emph{{\StrGobbleLeft{#1}{1}}}}

\usepackage[backend=bibtex,style=numeric-comp,sorting=none,giveninits=true,citestyle=numeric]{biblatex}
\usepackage[autostyle=true]{csquotes}

\title{
Investigating past high-energy fluxes with paleo-detectors
}

\author{
	\authorstyle{C.~Galelli\textcolor{GammaDarkGreen}{\textsuperscript{1}}}
	\newline\newline
    {\footnotesize
        \textsuperscript{\color{GammaDarkGreen}1}\institution{INFN Milano, via Celoria 16, 20133, Milano, Italy.}
    }\\
}

\date{\today}
\newcommand{\website}{\href{https://plan.events.mpg.de/event/543/}{Proceedings of the 9th Heidelberg Gamma-Ray Symposium}} 

\begin{document}
\sansmath\maketitle
\thispagestyle{fancy}

\lettrineabstract{Paleo-detectors provide a unique avenue to reconstruct the multi-million-year history of cosmic-ray (CR) flux, preserving signatures of transient high-energy events such as nearby supernovae. This technique aims to use natural minerals as particle detectors, looking at the persistent damage tracks created by CR-induced nuclear recoils, accumulated over the minerals’ geological lifespan, offering a geological archive of past particle fluxes. Building on our simulation study of Messinian Salinity Crisis evaporites, which demonstrated that minerals with specific geological histories may enable the detection of primary CR flux variations, we have now expanded to diverse terrestrial records. This contribution presents our recent publication proposing olivine xenoliths from Auvergne, France, where eruption chronosequences could allow for the differentiation of CR flux scenarios over the last 50 kyr. This phenomenological work is supported by the INFN-funded PRImuS experiment, which aims to prove the efficacy of high-throughput optical microscopy and plasma etching to analyze these mineral targets. By validating theoretical track-length spectra and refining the estimates of experimental effects, PRImuS’s goal is to establish paleo-detectors as a powerful tool for very-long-range time-domain astrophysics.}

\section{Introduction}
\label{sec:intro}
 
Cosmic rays are highly energetic particles, predominantly protons ($\sim 87\%$), helium nuclei ($\sim 12\%$), and a small admixture of heavier nuclei, accelerated in some of the most extreme environments in the Universe and spanning an enormous range of energy and flux. During propagation from their sources to Earth, charged cosmic rays are deflected and delayed by Galactic magnetic fields, so that their sources and time of emission cannot be reconstructed with good precision directly from present-day arrival directions. Direct flux measurements cover less than a century, and we can only observe indirect echoes of past acceleration episodes like nearby supernovae, supernova remnants, gamma-ray bursts, and AGN activity.
 
Cosmogenic-isotope proxies ($^{14}$C, $^{10}$Be, $^{26}$Al) extend the reachable window to $\mathcal{O}(10^5)$~yr, and in some cases provide good time resolution, but they are entangled with systematics from the
geomagnetic field strength and from carbon-cycle and climate variability. Independent hints that the CR flux has varied on longer timescales come from proposed correlations between mass-extinction events and hypothesized nearby-supernova CR spikes. What is missing is an archive that passively freezes the passage of particles, in the same way a sediment preserves a fossil.
 
\begin{figure}[t]
  \centering
  \includegraphics[width=0.4\textwidth]{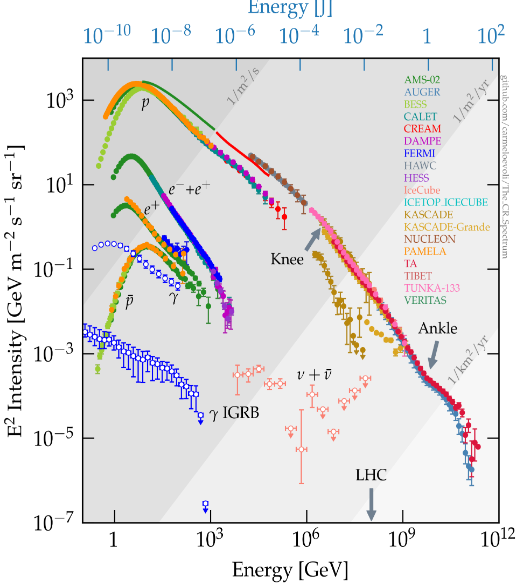}
  \caption{The all-particle cosmic-ray spectrum from GeV to beyond $10^{11}$~GeV, compiled from multiple experiments, illustrating the enormous dynamic range in energy and flux.}
  \label{fig:crspectrum}
\end{figure}
 
\section{Paleo-detectors as archives of nuclear recoils}
\label{sec:primer}
 
The paleo-detector technique proposes to use natural minerals as solid-state track detectors accumulating damage over geological timescales \cite{Baum:2023whitepaper}. Energetic nuclear recoils inside a crystal
displace atoms from the lattice, leaving chains of atomic vacancies that form latent tracks of nanometre-to-micrometre length. This is precisely the physical basis of fission-track and alpha-recoil-track dating, both
long-established tools in geochronology. In the context of cosmic rays, a secondary CR-induced nuclear recoil dislodges a nucleus from the lattice and, in some cases, transforms it into a cosmogenic isotope, leaving behind both a track and a distinctive isotopic signature. Crucially, once formed, these tracks can survive for kyr to even Gyr timescales as long as no subsequent thermal or pressure annealing resets the crystal.
 
The paleo-detector approach was originally developed to exploit the enormous integrated exposure (mass $\times$ time) offered by natural samples in searches for rare interactions such as WIMP dark matter and
coherent elastic neutrino-nucleus scattering \cite{Baum:2023whitepaper}. In that context, the cosmic-ray-induced nuclear recoil rate is itself an important, and often dominant, background that must be suppressed by selecting samples with low intrinsic radioactivity and a well-understood shielding history. Here we turn the
argument around: because every mineral is exposed to the CR flux with an intensity that depends on depth and geological history, the background becomes the signal of interest, and essentially any mineral with a well-constrained exposure history can be a candidate paleo-detector for cosmic rays.
 
\begin{figure}[t]
  \centering
  \includegraphics[width=0.45\textwidth]{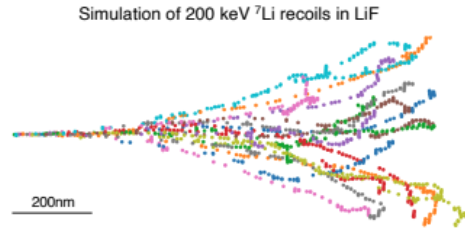}
  \caption{Simulated $200$~keV $^{7}$Li recoil tracks in LiF (P. Huber \&
  S. Hedges, 2026 MDvDM workshop, proceedings in preparation).}
  \label{fig:trackformation}
\end{figure}

\begin{figure}[t]
  \centering
  \includegraphics[width=0.45\textwidth]{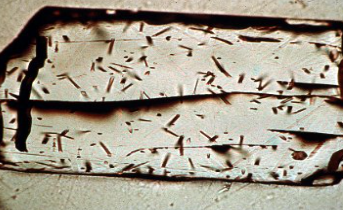}
  \caption{A microscope photograph showing fission tracks in apatite.}
  \label{fig:trackformation}
\end{figure}
 
\section{From primary cosmic rays to track-length spectra}
\label{sec:pipeline}
 
Understanding quantitatively the possible track signal contained in a candidate mineral requires a simulation pipeline connecting the primary CR spectrum to an observable track-length distribution. We identify three necessary ingredients:
\begin{enumerate}[label=\color{GammaRed}\large\theenumi] 
\item flux variations, i.e. the effect on the CR spectrum of transient or persistent astrophysical events, propagated down to the secondary muon and neutron spectra at Earth; 
\item interactions, i.e. the
propagation of secondary cosmic rays through the target material and the resulting energy distribution of recoiling nuclei; 
\item recoil length, i.e. the conversion of a given recoil energy into an expected
track length for each atomic species in the target.
\end{enumerate}
 
These three tasks are handled respectively by a custom MCEq \cite{Fedynitch:2015mcq} wrapper, which evolves the primary cosmic-ray spectrum into the cascade of secondary muons and neutrons; by a dedicated Geant4 \cite{Agostinelli:2003geant4} simulation, which propagates these secondaries through an infinite cylinder of the chosen target material and records, for every interaction, the depth, the recoiling species, the recoil energy and the residual energy of the incident particle; and by SRIM \cite{Ziegler:2010srim}, which provides the stopping range of recoiling ions in matter and is used to convert the recoil-energy spectrum $dR/dE$ into a track-length spectrum $dR/dx$. The three codes are combined, together with the target mass and the exposure time, into an integrated number of tracks per unit track length; the full pipeline is publicly available at \url{https://github.com/cgalelli/PrimusCode} \cite{PrimusCode}.
 
Because the astrophysical history of a sample is rarely a simple constant flux, the code separates two independent components. A \texttt{FluxHistory} object superimposes a baseline MCEq flux with any number of discrete or transient events (e.g., a nearby supernova), each cached independently; this part of the code is fully decoupled from the target-specific simulation. An \texttt{overburden\_history} object instead describes the burial history of the sample as a sequence of continuous deposition segments and discrete burial events, each with its own rate and density. At every timestep, the instantaneous flux is convolved with the Geant4 recoil-energy profiles appropriate for the overburden at that time, weighted by the interaction probability, and the resulting recoil rate is convolved with the SRIM range tables to obtain the differential track-production rate.
 
\section{Case study I: Messinian halite}
\label{sec:messinian}
 
As a first application, we consider halite (NaCl) deposits formed during the Messinian salinity crisis (5.97--5.33~Ma), when the near-total desiccation of the Mediterranean Sea produced massive evaporite deposits. These minerals offer a well-defined CR exposure window of $\sim 500$~kyr: they formed and remained exposed near the surface or under a shallow water column before being rapidly buried under kilometres of water and sediment at the Zanclean flood. Evaporites were deposited continuously throughout the crisis, with deposition-rate models in the literature ranging between $2.5$ and $30$~m/kyr, and marine evaporites are expected to carry very low intrinsic radioactivity ($\mathrm{U}\sim 10^{-8}$~g/g), keeping the radiogenic background well under control.
 
\begin{figure}[t]
  \centering
  \includegraphics[width=0.45\textwidth]{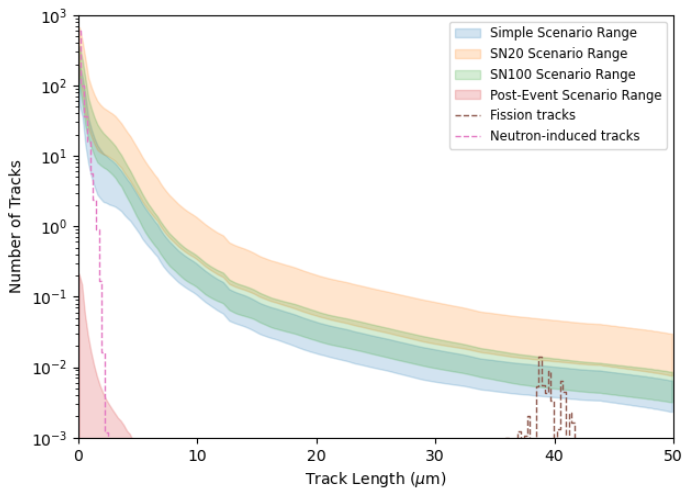}
  \caption{Predicted track-length spectra in Messinian halite for a simple flux scenario and for scenarios including a supernova at $20$ and $100$~pc, together with the fission-track and neutron-induced-track backgrounds; bands reflect the range of allowed deposition rates.
  }
  \label{fig:messinian}
\end{figure}

We compared three CR flux scenarios a simple, present-day-like flux, and two scenarios including a recent (at time of crystal formation) supernova at $20$ and $100$~pc, respectively, propagated through the full range of allowed deposition rates. The resulting track-length spectra form density bands that overlap but remain distinguishable across the three scenarios (Fig.~\ref{fig:messinian}), showing that the modeled enhancement of the primary flux from a nearby supernova leaves a detectable imprint even after marginalising over deposition-rate uncertainties. Tracks accumulated after burial, once the km-scale overburden has suppressed the secondary CR flux to a negligible level, are essentially absent, confirming that the
exposure window is cleanly bounded. This analysis has been published in \cite{Caccianiga:2024mess}.
 
\section{Case study II: a volcanic chronosequence}
\label{sec:volcanic}
 
A second, complementary case study exploits the Cha\^ine des Puys, a volcanic region in the Auvergne (France) where a large number of eruptions built the present-day landscape between roughly $100$ and $7$~kyr ago. Crucially, most of these eruptions sourced olivine-bearing xenoliths (peridotite) from the same magmatic environment in the upper mantle. Since individual eruptions bring these xenoliths to the surface, and hence into the CR flux, at well-dated times, samples from successive eruptions along the chain constitute a natural chronosequence: a time-resolved array of paleo-detectors sharing the same mineralogy and pre-eruption (CR-shielded) history, but differing systematically in their surface exposure time.
 
\begin{figure}[t]
  \centering
  \includegraphics[width=0.45\textwidth]{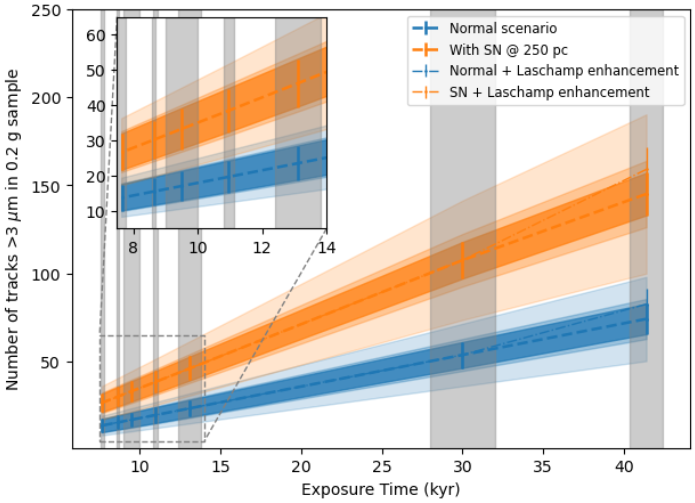}
  \caption{Expected number of tracks in a $0.2$~g olivine sample as a function of exposure time, for a normal-flux scenario, a scenario including the Antlia supernova precursor at  $250$~pc, and both combined with the Laschamp excursion enhancement; grey bands mark the ages of the sampled eruptions.}
  \label{fig:volcanic}
\end{figure}

We modelled the expected track yield in olivine as a function of exposure time against a toy Antlia supernova-remnant precursor at $250$~pc and against the Laschamp geomagnetic excursion ($\sim 41$~kyr ago, during which
the field dropped to roughly one sixth of its present strength and reversed). The normal-flux and supernova-enhanced scenarios are separated at up to the $1\sigma$ level once a conservative $30\%$ counting systematic is included, while the enhancement expected during the Laschamp excursion is separately detectable at $1\sigma$ with a more optimistic $10\%$ systematic in the no-supernova baseline (Fig.~\ref{fig:volcanic}). This result establishes, for the first time, a phenomenological path towards a time-resolved paleo-detector measurement of the CR flux, and has been published in \cite{Galelli:2026volc}.
 
\section{PRImuS: an experimental proof of concept}
\label{sec:primus}
 
The phenomenological studies above motivate an experimental proof of concept, PRImuS (Paleo-astroparticles Reconstructed with the Interactions of MUons in Stone), funded through an INFN grant for young researchers
(2025-2026, PI C.\ Galelli). PRImuS targets two complementary technical goals. The first is the development of argon plasma etching protocols as an alternative to conventional wet chemical etching: plasma etching is expected to offer a more efficient, eco-friendly, and safer route to revealing tracks, and because it relies on a purely kinetic, chemically inert process should provide a more faithful enlargement of tracks with less deformation of their morphology than aggressive chemical etchants such as hydrofluoric acid. At present, a stable etching recipe has been established for halite, with work ongoing to extend the protocol
to harder target minerals.
 
The second goal is an established high-throughput optical microscopy workflow, needed both because the size of the expected signal (etch pits of order microns) is optimally matched to optical resolution, and because scanning a statistically meaningful sample area requires automation of the stage, autofocus, and image-acquisition chain. The current end-to-end workspace throughput is of order $5$~cm$^2$ scanned at $20$ $z$-slices per day.
 
\section{OptimusPrimus: from images to spectra}
\label{sec:optimusprimus}
 
Converting raw microscope images into a track-length spectrum that can be compared to the theoretical predictions of Sec.~\ref{sec:pipeline} requires a dedicated image-analysis pipeline. The \texttt{OptimusPrimus} pipeline is built around a dual PyTorch U-Net architecture: a first semantic-segmentation network (an MAnet decoder with an EfficientNet-B7 encoder) is tuned for high recall in generating candidate track pixels across a stack of $z$-slices, producing a cumulative per-pixel signal probability; a probability threshold defines a binary candidate mask, from which connected components are extracted as fixed-size patches, geometrically normalised by ellipse fitting and major-axis alignment. A second, dedicated classification network then refines these candidates for high-precision track/background discrimination. The network is trained and validated on obsidian fission tracks, and the current aim is to surpass $85\%$ recall and $75\%$ precision in identification.
 
\begin{figure}[t]
  \centering
  \includegraphics[width=0.45\textwidth]{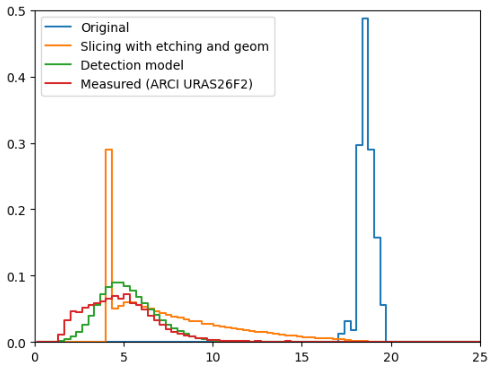}
  \caption{Comparison of the original 3D theoretical spectrum, its projection after simulating slicing, etching, and detection effects, and the measured spectrum in an obsidian reference sample.}
  \label{fig:optimusprimus}
\end{figure}

Because the theoretical spectrum is naturally expressed as a volumetric quantity (tracks per unit mass), while the microscope measures a two-dimensional projection (tracks per unit area on a given cleave or polished surface), a dedicated algorithm bridges the two. Starting from a Monte Carlo realisation of the expected 3D track population, with a random starting point and orientation, the module computes each track's endpoint, selects a slicing plane, retains only the tracks intersecting that plane, projects them onto it, and finally ``measures'' the projected tracks accounting for the finite size of the etch pit. An additional step folds in the identification efficiency of the OptimusPrimus network itself, using its measured recall and precision. This procedure has been validated against real track-length distributions measured in an obsidian reference sample (ARCI URAS26F2; Fig.~\ref{fig:optimusprimus}), showing good agreement once slicing, etching, and detection effects are properly propagated.
 
\section{Status and outlook}
\label{sec:outlook}
 
PRImuS will run until December 2026; the current priority is to analyse as much material as possible before drawing quantitative conclusions. To date, $\mathcal{O}(50)$~g of xenoliths have been collected from the Puy de Montcineyre eruption ($\sim 7$~kyr). PRImuS aims to provide a solid proof of concept, but scaling this technique into a genuine measurement will require larger and more diverse sample sets and, ideally, a readout
technique that is faster and less destructive than etching alone.
 
This last point motivates our planned follow-up project. In many candidate minerals, including halite and
olivine, electrons trapped in the vacancies that compose a latent track become optically active \emph{colour centres} and fluoresce under suitable illumination. We plan to exploit the mesoSPIM platform, a low-cost,
open-source light-sheet fluorescence microscope originally developed for imaging cleared brains and organs in neuroscience (\url{https://mesospim.org/}) \cite{mesoSPIM, Vladimirov:2024meso}, which allows fast, non-destructive three-dimensional bulk imaging by scanning large volumes as a sequence of 2D fluorescence slices; the technique has already been verified on
irradiated LiF samples. The GALACTIC strategy is to image bulk colour centres first (fast and non-destructive), and only then selectively etch the same sample to reveal tracks (destructive, following the PRImuS baseline), cross-calibrating the two observables on a common sample set.
 
\section{Conclusions}
\label{sec:conclusions}
 
Reconstructing the past evolution of the cosmic-ray flux would benefit not only astrophysics, but also particle physics, geophysics, and potentially even paleontology and archaeology, wherever a mineral's exposure history can be used to date or cross-check other geochronological indicators. We have shown that natural minerals with well-constrained burial histories could be sensitive to astrophysical and geomagnetic events shaping the CR spectrum at Earth, opening a genuinely new window on long-timescale, time-domain high-energy astrophysics. PRImuS constitutes a first proof of concept for the experimental side of this programme, and, together with the phenomenological simulation, the results motivate the design of a dedicated cosmic-ray paleo-detection experiment.

\printbibliography[title={References}]

@article{Baum:2023whitepaper,
    author =       "Baum, Sebastian and Stengel, Patrick and Abe, Natsue
                     and Acevedo, Javier F. and Araujo, Gabriela R. and
                     Asahara, Yoshihiro and Avignone, Frank and Balogh,
                     Levente and Baudis, Laura and Boukhtouchen, Yilda
                     and Bramante, Joseph and Breur, Pieter Alexander
                     and Caccianiga, Lorenzo and Capozzi, Francesco and
                     Collar, Juan I. and Ebadi, Reza and Edwards, Thomas
                     and Eitel, Klaus and Elykov, Alexey and Ewing,
                     Rodney C. and Freese, Katherine and Fung, Audrey
                     and Galelli, Claudio and Glasmacher, Ulrich A. and
                     Gleason, Arianna and Hasebe, Noriko and Hirose,
                     Shigenobu and Horiuchi, Shunsaku and Hoshino,
                     Yasushi and Huber, Patrick and Ido, Yuki and Igami,
                     Yohei and Itow, Yoshitaka and Kato, Takenori and
                     Kavanagh, Bradley J. and Kawamura, Yoji and Kazama,
                     Shingo and Kenney, Christopher J. and Kilminster,
                     Ben and Kouketsu, Yui and Kozaka, Yukiko and
                     Kurinsky, Noah A. and Leybourne, Matthew and Lucas,
                     Thalles and McDonough, William F. and Marshall,
                     Mason C. and Mateos, Jose Maria and Mathur, Anubhav
                     and Michibayashi, Katsuyoshi and Mkhonto, Sharlotte
                     and Murase, Kohta and Naka, Tatsuhiro and Oguni,
                     Kenji and Rajendran, Surjeet and Sakane, Hitoshi
                     and Sala, Paola and Scholberg, Kate and Semenec,
                     Ingrida and Shiraishi, Takuya and Spitz, Joshua and
                     Sun, Kai and Suzuki, Katsuhiko and Tanin, Erwin H.
                     and Vincent, Aaron and Vladimirov, Nikita and
                     Walsworth, Ronald L. and Watanabe, Hiroko",
    title =        "{Mineral detection of neutrinos and dark matter. A
                     Whitepaper}",
    year =         "2023",
    month =        "1",
    archivePrefix = "arXiv"
}

@article{Caccianiga:2024mess,
    author =       "Caccianiga, Lorenzo and Apollonio, Lorenzo and
                     Mariani, Federico Maria and Magnani, Paolo and
                     Galelli, Claudio and Veutro, Alessandro",
    title =        "{Sedimentary rocks from Mediterranean drought in the
                     Messinian age as a probe of the past cosmic ray flux}",
    journal =      "Phys. Rev. D",
    volume =       "110",
    pages =        "L121301",
    year =         "2024",
    DOI =          "10.1103/PhysRevD.110.L121301"
}

@article{Galelli:2026volc,
    author =       "Galelli, Claudio and Caccianiga, Lorenzo and
                     Apollonio, Lorenzo and Magnani, Paolo and Breton,
                     Vincent",
    title =        "{A volcanic chronosequence as a time-resolved
                     paleo-detector array to study the cosmic-ray flux
                     in the late Pleistocene and Holocene}",
    journal =      "JCAP",
    volume =       "04",
    pages =        "023",
    year =         "2026",
    DOI =          "10.1088/1475-7516/2026/04/023"
}

@article{Fedynitch:2015mcq,
    author =       "Fedynitch, Anatoli and Engel, Ralph and Gaisser,
                     Thomas K. and Riehn, Felix and Stanev, Todor",
    title =        "{Calculation of conventional and prompt lepton
                     fluxes at very high energy}",
    journal =      "EPJ Web Conf.",
    volume =       "99",
    pages =        "08001",
    year =         "2015",
    DOI =          "10.1051/epjconf/20159908001"
}

@article{Agostinelli:2003geant4,
    author =       "Agostinelli, S. and others",
    title =        "{GEANT4 -- a simulation toolkit}",
    journal =      "Nucl. Instrum. Meth. A",
    volume =       "506",
    pages =        "250--303",
    year =         "2003",
    DOI =          "10.1016/S0168-9002(03)01368-8"
}

@article{Ziegler:2010srim,
    author =       "Ziegler, James F. and Ziegler, M. D. and Biersack,
                     J. P.",
    title =        "{SRIM -- The stopping and range of ions in matter
                     (2010)}",
    journal =      "Nucl. Instrum. Meth. B",
    volume =       "268",
    number =       "11-12",
    pages =        "1818--1823",
    year =         "2010",
    DOI =          "10.1016/j.nimb.2010.02.091"
}

@misc{PrimusCode,
    author =       "{Galelli, C.}",
    title =        "{PrimusCode: simulation pipeline for cosmic-ray
                     paleo-detectors}",
    howpublished = "\url{https://github.com/cgalelli/PrimusCode}"
}

@misc{mesoSPIM,
    author =       "{The mesoSPIM initiative}",
    title =        "{mesoSPIM: an open-source light-sheet microscope
                     for imaging cleared tissue}",
    howpublished = "\url{https://mesospim.org/}"
}

@article{Vladimirov:2024meso,
    author = "{Vladimirov, N., et al}",
    title = "{Benchtop mesoSPIM: a next-generation open-source light-sheet microscope for cleared samples.}",
    journal = "{Nature Communications}",
    year = "{2024}",
    doi = "{10.1038/s41467-024-46770-2}"
}

\end{document}